\documentclass{ifacconf}
\usepackage{graphicx}      % include this line if your document contains figures
\usepackage{natbib}        % required for bibliography
\usepackage{amsmath} % assumes amsmath package installed
\usepackage{amssymb}  % assumes amsmath package installed
\usepackage{bm}
\usepackage{algorithmic}
\usepackage{algorithm} 
\usepackage{blindtext}
\usepackage{multirow}
\usepackage{color}
\usepackage{soul}
\usepackage{url} 
\usepackage{subfigure}

\newcommand{\layerindex}{l}
\def\rmI{{\mathbf{I}}}

\newcommand{\bz}{\mathbf{z}}

\newcommand{\R}{\mathbb{R}}

\newcommand{\bN}{\mathcal{N}}

\newcommand{\define}{\triangleq}

\newcommand{\by}{\mathbf{y}}
\newcommand{\bw}{W}
\newcommand{\btheta}{{\sigma^2}}
\newcommand{\Prob}{p}

\newcommand{\hyper}{\upsilon}
\newcommand{\Hyper}{\boldsymbol{\upsilon}}
\newcommand{\net}{\mathcal{W}}
\newcommand{\prior}{p}
\newcommand{\bG}{\Upsilon}
\newcommand{\bGamma}{\Upsilon}

\newcommand{\bgamma}{\boldsymbol{\upsilon}}
\newcommand{\br}{\upsilon}

\newcommand{\Hessian}{{\mathbf{H}}}
\newcommand{\Gradient}{{\mathbf{G}}}
\newcommand{\argmin}{\text{arg}\min}
\newcommand{\mean}{\mathbf{\mu}}
\newcommand{\balpha}{\boldsymbol{\alpha}}
\newcommand{\quadratic}{{\frac{1}{2}(\bw^l-{\bw^l}^*)^{\top}\Hessian({\bw^l}^{*}, \btheta)(\bw^l-{\bw^l}^*)+(\bw^l-{\bw^l}^*)^{\top}\mathbf{G}({\bw^l}^*, \btheta)}}

\newcommand{\energy}{\frac{1}{2\sigma^2}\sum_{t=1}^{N}(y(t)-\text{Net}(\bz(t),\net))^2}
\newcommand{\normone}{\ell_1}

\DeclareMathOperator{\diag}{diag}
\DeclareMathOperator*{\argmax}{argmax}

\begin{document}
\begin{sloppypar}
\begin{frontmatter}

\title{
A Sparse Bayesin Deep Learning Approach for Identificaiton of Cascaded Tanks Benchmark
\thanksref{footnoteinfo}}

\thanks[footnoteinfo]{The work of Hongpeng Zhou is sponsored by the program of China Scholarships Council (No.201706120017). The authors are with the Faculty of Mechanical, Maritime and Materials Engineering, Delft University of Technology, 2628 CD Delft, The Netherlands. 
For correspondence: {\tt\small wei.pan@tudelft.nl}.
}

\author[First]{Hongpeng Zhou} 
\author[Second]{Chahine Ibrahim} 
\author[Third]{Wei Pan}

\address[First]{Department of Cognitive Robotics, TU Delft}
\address[Second]{Delft Center for Systems and Control, TU Delft}
\address[Third]{Department of Cognitive Robotics, TU Delft}

\begin{abstract}                % Abstract of not more than 250 words.
Nonlinear system identification is important with 
a wide range of applications. 
The typical approaches for nonlinear system identification include Volterra series models, nonlinear autoregressive with exogenous inputs models, block-structured models, state-space models and neural network models. Among them, neural networks (NN) is an important black-box method thanks to its universal approximation capability and less dependency on prior information. 
However, there are several challenges associated with NN. The first one lies in the design of a proper neural network structure. A relatively simple network cannot approximate the feature of the system, while a complex model may lead to overfitting. 
The second lies in the availability of data for some nonlinear systems. For some systems, it is difficult to collect enough data to train a neural network. 
This raises the challenge that how to train a neural network for system identification with a small dataset. In addition, if the uncertainty of the NN parameter could be obtained, it would be also beneficial for further analysis. 
In this paper, we propose a sparse Bayesian deep learning approach to address the above problems. Specifically, the Bayesian method can reinforce the regularization on neural networks by introducing introduced sparsity-inducing priors. 
The Bayesian method can also compute the uncertainty of the NN parameter. An efficient iterative re-weighted algorithm is presented in this paper. 
We also test the capacity of our method to identify the system on various ratios of the original dataset. The one-step-ahead prediction experiment on Cascaded Tank System shows the effectiveness of our method. 
Furthermore, we test our algorithm with more challenging simulation experiment on this benchmark, which also outperforms other methods.
\end{abstract}
\end{frontmatter}
%===============================================================================
\section{Introduction}
Neural network (NN) is an important nonlinear system identification method. Its main advantages lie in two aspects. 
First of all, neural networks have the capability for universal approximation~\citep{Sontag1993SomeTI,  hornik1989multilayer}. They can approximate a nonlinear system, e.g. using Fully-Connected (FC) networks with several hidden layers for both static and dynamic nonlinear systems~\citep{narendra1992neural, narendra1990identification}; using recurrent neural networks for the dynamic identification and control of nonlinear systems~\citep{1643442, dinh2010dynamic}. Secondly, as a black-box method, a neural network could be trained on observed data without any prior information about the system~\citep{liu2012nonlinear}. 

However, there are some challenges in using neural networks for nonlinear system identification. The first challenge is the design of an appropriate neural network structure. A system cannot be approximated with a simple structure, while a complex one may lead to overfitting due to the intrinsic connections. Therefore, the balance between complexity and model prediction accuracy of neural network should be considered.
A second challenge lies in the availability of data. For some nonlinear systems, the collection of data is very difficult. In other words, a method that can train a neural network with small dataset is highly desired.

Given these problems, we notice that Bayesian system identification provides a probabilistic perspective in understanding and tackling these issues. Specifically, Bayesian approaches have the following advantages:
1) Overfitting could be alleviated by penalizing over model parameters~\citep{mackay1995probable};
2) The uncertainty of the model could be quantified, which would be beneficial for analysis~\citep{Peterka:1981:BSI:2233038.2233131};
3) Fewer tuning of hyper-parameters ~\citep{mackay1992Bayesian,hernandez2015probabilistic};
4) Sparsity-inducing hyper-priors eliminates model redundancies ~\citep{pan2016sparse, 8307489,pan2017Bayesian}. 
A range of Bayesian nonlinear identification solutions have been developed in the last decades. To name just a few examples,
\cite{pan2016sparse,pan2014distributed,pan2017Bayesian} proposed a sparse Bayesian approach for nonlinear state-space systems. By combining Gaussian process regression, Bayesian nonparametric methods are developed for NARX modelling~\citep{DBLP:journals/corr/abs-1303-2912,doi:10.1080/13873950500068567}.
It was also explored to identify biochemical reaction networks from single dataset~\citep{pan2012reconstruction}and heterogeneous datasets~\citep{pan2015identifying}, respectively. And recently a cyber-physical system modelled by hybrid system can be automatically discovered from data \citep{yuan2019data}. Furthermore, the sparse Bayesian learning algorithms using automatic relevance determination (ARD) also have been applied for different applications, e.g. regression and classification~\citep{DBLP:journals/corr/abs-1301-3838, tipping2001sparse}, signal denoising~\citep{Zhang2008SignalDA}, neural architecture search~\citep{zhou2019bayesnas} and pattern recognition~\citep{BayesRecognition}.

Inspired by the advantage of Bayesian method, we propose to address the challenges of neural network models from two aspects: a) the neural network could be modelled from Bayesian perspective; b) a sparse neural network could be obtained by applying a sparse Bayesian learning approach. Specifically, the objective is to maximize the posterior estimation of the weight matrices. To compute intractable marginal likelihood, we adopt the Laplace approximation method. The likelihood is assumed to be Gaussian distribution. 
The Gaussian priors for the weight matrices are incorporated to promote sparsity. 
Under these conditions, the Bayesian nonlinear system identification could be regarded as a type II maximum likelihood problem~\citep{tipping2001sparse}. An re-weighted algorithm is proposed to estimate the system parameter and introduced hyper-parameter iteratively. 

The proposed method could reduce the burden on network design for different applications. We could initialize a deep and complex network structure to guarantee the ability to approximate the nonlinear system. And a very sparse Bayesian neural network could be obtained in the end which will address the overfitting problem. At the same time, the relative sparse structure also reduces the requirement for huge data. Two experiments are implemented on Cascaded tank system~\citep{cascad_benchmark} in this paper. The first one is the one-step ahead prediction experiment which shows the proposed approach could keep the prediction performance even with a small dataset. In addition, we also did the simulation experiment, which is more challenging. The simulation result further support that our method could learn the intrinsic feature of a system or process.

The paper is organized as follows. In Section~\ref{sec:problem formulation},  the general nonlinear system identification problem is formulated by using neural network. The nonlinear system identification problem is formulated from the perspective of the Bayesian theory in Section~\ref{sec:Bayesian theory}. In Section~\ref{sec:proposed algorithm}, the proposed sparse Bayesian approach is elaborated. Details about both the prediction and simulation experiments can be found in Section~\ref{sec:experiment}. In Section~\ref{sec:conclusion}, we conclude and discuss several problems that we plan to study in the future. 
% The details derivation procedures and proof content for the proposed method are in Appendix~\ref{app:proof for proposition 1},\ref{app:proof for proposition 2},\ref{appsec:cccp_procedure}.

\section{Problem Formulation}
\label{sec:problem formulation}
The objective of the neural network approach is to approximate the system input/output relations based on a finite dataset. 
Given a nonlinear system with collected regressor vector $\bz(t) = [u(t),u(t-1), \ldots ,u(t-n_a),y(t-1),\ldots ,y(t-{n_b})]\in R^{n_a+n_b+1}$, where $u(t),y(t)$ stands for the system input and output and $n_a, n_b$ are two scalars to define the maximum lags. For simplicity, we assume both $u(t)$ and $y(t)$ be scalars, and the identification problem with a neural network could be formulated as building following mathematical model:
\begin{equation}
\hat{y}(t+1)= \text{Net}(\bz(t),\net(t)) +\xi 
\label{eq:model_formulation}
\end{equation}
where $\xi$ is the noise with zero-mean normally distribution $\xi\thicksim\bN(0,\sigma^2)$. 
$\text{Net}(\cdot)$ represents the nonlinear mapping $R^{n_a+n_b}\rightarrow R$, which will be realized by the neural network. 
$\net(t)$ denotes the weight matrices to be optimized in the neural network. If we suppose the neural network includes $L$ layers, $\net(t)$ could be denoted as \begin{math}
\net(t) \triangleq \left[\bw(t)^1,\bw(t)^2,\ldots,\bw(t)^L\right]
\end{math}. And 
\begin{math}
\bw(t)^l\in\mathbb{R}^{n_{l-1}\times n_l} (l=1,\ldots,L)
\end{math} stands for the weight matrix between layer $l-1$ and layer $l$, where $n_{l-1}$ and $n_l$ are the number of neurons in these two layers.
\begin{figure}[t]
	\begin{center}
		\includegraphics[scale = 0.25]{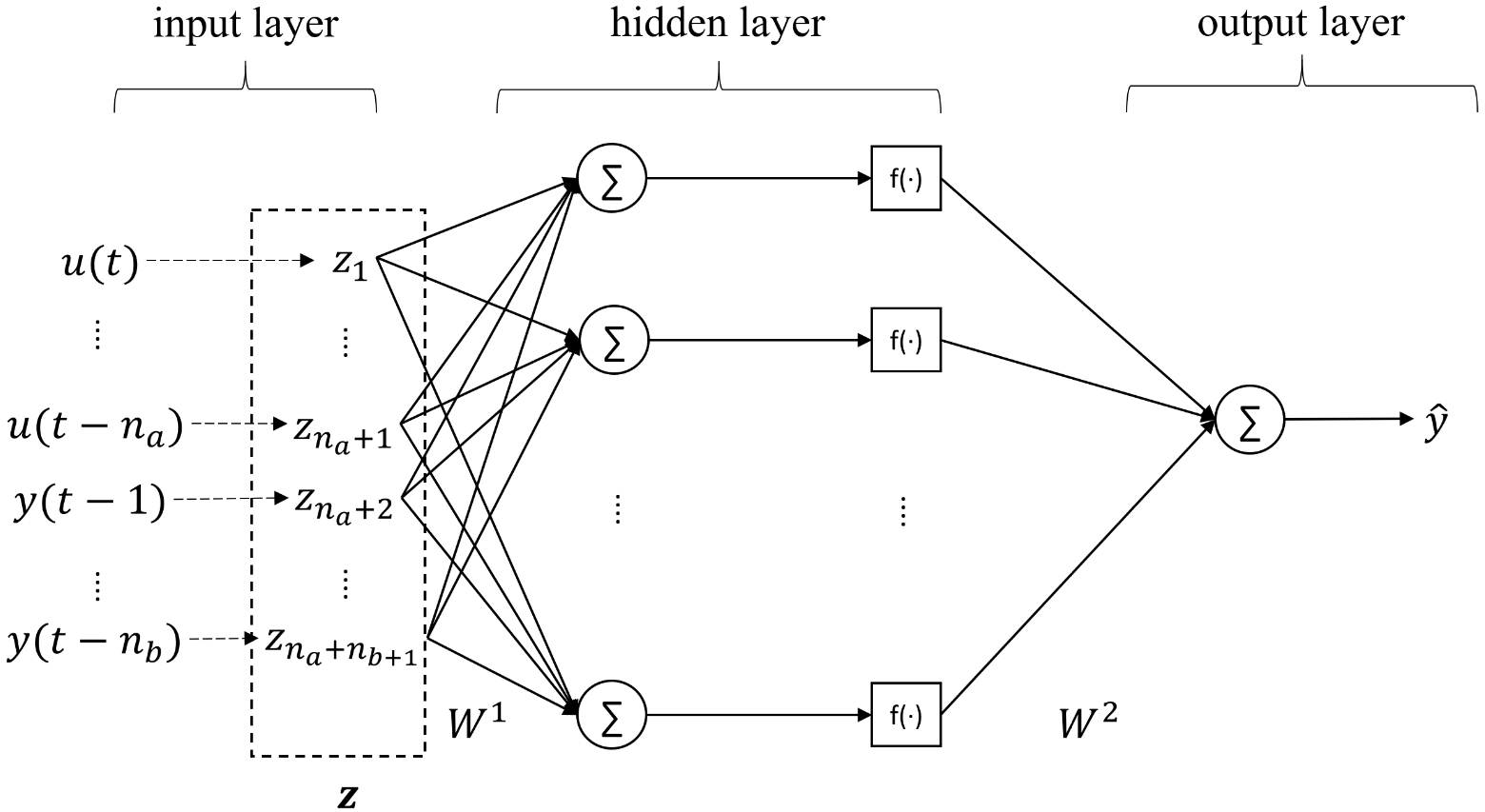}
		\caption{A multi-layer perceptron with one hidden layer.} 
		\label{fig:mlp_network}
	\end{center}
\end{figure}

There are many different typical structures for a neural network, e.g. Convolutional Neural Network, Recurrent Neural Network and Multi-Layer Perceptron (MLP). MLP is the most common architecture, which is a feed-forward network constructed by arranging perceptron-type neurons in different layers. In the following, we will take MLP as an example to explain how the neural network works for system identification. And Fig.~\ref{fig:mlp_network} is a schematic representation of a MLP with a single hidden layer. 

As a fully connected network, every neuron in MLP is connected to other neurons in the adjacent layers. As shown in Fig.~\ref{fig:mlp_network}, $a_{j}^1$, the output of the \textit{j-th} neuron in the hidden layer, could be computed as:
\begin{subequations}
\begin{align}
h_{j}^1 = {\bw^1}\bz + b^1 &= \sum\limits_{i=1}^{n_a+n_b+1} ({\bw_{ij}^1}{z_{i}} + b_j^1), 
\label{eq:single_neuron_combination} 
\\
a_{j}^1 &= f(h_{j}^1)
\label{eq:single_neuron_activation}
\end{align}
\end{subequations}
where $\bz$ is the input of the network; $\bw_{ij}^1$ is the weighted scalar which determines the strength of the connection between the \textit{i-th} neuron in the input layer and \textit{j-th} neuron in the hidden layer. $b$ is the bias. Eq.~(\ref{eq:single_neuron_combination}) stands for the linear combination, which is computed as the sum of product between the weight matrix $\bw^1$ and input vector $\bz$.
The nonlinear behaviour of the neural network model is decided by $f(\cdot)$, which is named as the nonlinear activation function. The common activation functions include the logistic sigmoid function \begin{math} a = \frac{1}{1+e^{-h}} \end{math}, the hyperbolic tangent function \begin{math} a = \frac{e^{2h}-1}{e^{2h}+1} \end{math} and the point-wise rectified linear units \begin{math} a = max(h,0) \end{math}.

By using the neural network for system identification, we need to find the optimal weight parameters for fitting the observations according to the criterion function, i.e. root mean square error (RMSE), i.e. \begin{math}
\text{RMSE}(\hat{y},y) = \sqrt{\frac{1}{N}\sum\limits_{t=1}^{N} ({y(t) - \hat{y}(t)})^2}
\end{math}, where $\hat{y}$ is the predicted output and $y$ is the true output.
With the defined criterion function, the training for a neural network could be implemented with stochastic gradient descent (SGD). The predicted errors assessed by the criterion function will be back-propagated through the network. And both the weights and biases of the network will be updated iteratively during the training process. However, the conventional method for training a neural network cannot solve the overfitting problem. A sparser and simpler network is expected to approximate the non-linear system accurately to a sufficient extent. In the next section, this problem will be addressed from a Bayesian perspective. 

\section{A Bayesian FRAMEWORK FOR NONLINEAR SYSTEM IDENTIFICATION}
\label{sec:Bayesian theory}
Bayesian method offers a principal way to infer certain probability distributions for the unknown parameters $\bw$. In this section, we will explain the nonlinear system identification problem within a Bayesian framework. For easy notation, $\bw$ in this section denotes $\bw(t)^l$ which represents any 2-D weight matrix in a neural network and has been defined in Section~\ref{sec:problem formulation}. 
\subsection{Bayesian Posterior Estimation}
Given the measured data set $\by=[y(1),...,y(N)]$, the posterior distribution for parameters $\bw$ could be denoted as:
\begin{equation}
\label{eq:general posterior distribution}
p(\bw|\by)= \frac{p(\by|\bw)p(\bw)}{p(\by)}
\end{equation}
where $p(\by|\bw)$ is the likelihood function and $p(\bw)$ refers to the prior distribution for $\bw$; the marginal likelihood $p(\by)$ could be calculated as 
\begin{math}
p(\by) = \int p(\by|\bw)p(\bw)d\bw
\end{math}.
It is obvious that to achieve the posterior distribution, we should calculate the items in both numerator and denominator in Eq.~\eqref{eq:general posterior distribution}.

\subsection{Likelihood Function}
We suppose the distribution of data likelihood belongs to a Normal distribution:
{
	\fontsize{9.5}{2}
\begin{equation}
\begin{aligned}
&p(\by|\bw,\sigma^2) = \prod_{t=1}^N \mathcal{N}(y(t)| \text{Net}(\bz(t),\net); \sigma^2) \\
& = (2\pi\sigma^2)^{\frac{N}{2}}\exp\left(-\frac{1}{2\sigma^2}{\sum_{t=1}^{N}\left(y(t)-\text{Net}\left(\bz(t),\net\right)\right)^2}\right)
\label{general-likelihood-expo}
\end{aligned}
\end{equation}
}
\subsection{Priors}
As the likelihood given by Eq.~\eqref{general-likelihood-expo} is a member of the exponential family, it is also better to define the prior as an exponential distribution which satisfies the conjugacy requirement \citep{bishop2006pattern}. 
We define a Gaussian prior distribution $\Prob(\bw)$, which could enforce sparsity on $\bw$:
\begin{equation}
\begin{aligned}
\Prob(\bw)& =\prod_{i=1}^{n_{l-1}}\prod_{j=1}^{n_l}\bN({\bw_{ij}}|0,\hyper_{ij})\prior(\hyper_{ij}) 
\label{eq:relaxed gaussian prior}
\end{aligned}
\end{equation}
    where $\hyper_{ij}$ is an unknown parameter which should also be inferred. Therefore, we introduce the hyper-prior $\prior(\hyper_{ij})$, which should be a non-negative function. Considering gamma priors have the effect of promoting sparsity~\citep{tipping2001sparse}, we choose gamma distribution as the hyper-prior,\textit{i.e.}, $\prior(\hyper_{ij}) = \text{Gamma}(\hyper_{ij}|a, b)$~\citep{berger2013statistical}. 
To make the prior flat, $a$ and $b$ is simply fixed to be zero~\citep{tipping2001sparse}.
If we define $\boldsymbol{\hyper} \define \left[\hyper_1, \ldots, \hyper_{j},\ldots,\hyper_{n_l}\right] \in \R^{({n_{l-1}\times n_l},1)}_{+}$, where $\hyper_{j} = \left[\hyper_{1j}, \ldots, \hyper_{n_{l-1}j}\right]$,  a joint distribution $\bw$ and ${\bgamma}$ could be formed as:
\begin{equation}
\begin{aligned}
p(\bw, \bgamma) 
&= \prod_{i=1}^{n_{l-1}}\prod_{j=1}^{n_l}\bN(\bw_{ij}|0,\br_{ij}){p}(\hyper_{ij})\\
&=\bN(\bw|\mathbf{0},\bG){p}(\bgamma),
\label{eq:super gaussian prior}
\end{aligned}
\end{equation}
where \begin{math}
	\bG = \diag\left[\Hyper \right]
\end{math}.

\subsection{Marginal Likelihood}
\label{subsec:marginal likelihood}
After introducing the $\bgamma$, the key question is to select appropriate $\hat{\bgamma}=\left[\hat{\hyper}_{1},\ldots, \hat{\hyper}_{n_l}\right]$ to make $\Prob(\bw|\by,\hat{\bgamma},\sigma^2)$ be a close relaxation to the posterior distribution $\Prob(\bw|\by)$. If the parameter of the likelihood distribution $\sigma$ is fixed, the optimal $\hat{\bgamma}$ could be decided by the principle of minimizing the misaligned probability:
\begin{equation}
\begin{aligned}
\hat{\bgamma} =  \argmin\limits_{\bgamma\geq\mathbf{0}} \int \Prob(\by|\bw, \sigma^2)\left(\big|\Prob(\bw)-\Prob(\bw,\bgamma)\big|\right)d\bw
\label{mass}
\end{aligned}
\end{equation}
As $\Prob(\bw,\bgamma)\le\Prob(\bw)$, the solution for (\ref{mass}) can be converted as type-II maximum likelihood \citep{tipping2001sparse}:
\begin{equation}
\begin{aligned}
\hat{\bgamma} = \argmax\limits_{\bgamma\geq\mathbf{0}} 
\int p({\by}|\bw,\sigma^2) \bN( \bw|\mathbf{0},\bG)\prior(\bgamma)d\bw
\label{eq:generalmarignal}
\end{aligned}
\end{equation}
This also means that the optimal hyperparameters $\hat{\bgamma}$ can be computed by maximizing the marginal likelihood. 
\section{SPARSE Bayesian DEEP LEARNING}
\label{sec:proposed algorithm}
As elaborated in Section~\ref{subsec:marginal likelihood}, the nonlinear system identification problem could be regarded as an optimization problem. In this section, we will propose the optimization procedures and derive an iterative algorithm. 
\subsection{Optimization Problem Formulation}
% Two propositions will be illustrated in the first.  
We formulate the optimization problem in proposition~\ref{general:proposition1:cost} .
\begin{prop}
	\label{general:proposition1:cost}
	With the likelihood in Eq.~\eqref{general-likelihood-expo} and the sparse inducing prior in Eq.~\eqref{eq:super gaussian prior}, the unknown parameter $\bw^l$, hyperparameter $\bgamma^l$ could be obtained by solving following approximated optimization problem:
	\begin{equation}
	\begin{aligned}
	\min_{ \bw^l, \bgamma^l, \btheta} \mathcal{L}(\bw^l, \bgamma^l, \btheta) 
	\notag
	\end{aligned}
	\end{equation}
	with 
	{
		\fontsize{8.5}{2}
	\begin{equation}
	\label{general:proposition1}
	\begin{split}
	&\mathcal{L}(\bw^l, \bgamma^l, \btheta) \\
	= &{\bw^l}^{\top} \left[\Hessian^l+{\bG^l}^{-1}\right] \bw^l+ 2 {\bw^l}^{\top} \left[\Gradient^l-\Hessian^l {\bw^l}^{*}\right] + \log |\bG^l| \\ + &\log |\Hessian^l+ {\bG^l}^{-1} |
	-N\log {{(2\pi\sigma^2)} \cdot b^l} -2 \sum_{i=1}^{n_{l-1}}\sum_{j=1}^{n_{l}} \log \prior(\hyper_{ij}^l) 	
	\end{split}
	\end{equation}
}where $b^l \define b({\bw^l}^*, \btheta)$ with:
{
	\fontsize{8.1}{2}
	\begin{equation*}
	\begin{split}
	\exp \left\{-\frac{1}{2}{{\bw^l}^*}^{\top} \Hessian^l {{\bw^l}^*} + {{\bw^l}^*}^{\top} \Gradient^l - \frac{1}{2\sigma^2}\sum_{t=1}^{N}(y(t)-\text{Net}(\bz(t),\net^*))^2\right\}
	\end{split}
	\end{equation*}
}for simplicity, we use $\Hessian^l$ and $\Gradient^l$ to denote $\Gradient({\bw^l}^*, \btheta)$ and $\Hessian({\bw^l}^*, \btheta)$ which are the gradient and Hessian of ${\bw^l}^*$. 
% The analytical expression of $\Hessian$ and $\Gradient$ is illustrated in Eq~\eqref{eq:fc_gradient} and Eq~\eqref{eq:fc_hessian}.
\end{prop}
As analyzed in Section~\ref{subsec:marginal likelihood}, the key question is to select optimal $\hat{\bgamma}$ and then update $\bw$. Therefore, if we first initialize $\btheta$ to some reasonable value, the proposition~\ref{general:proposition1:cost} could be re-formulated as: 
\begin{prop}
	\label{general:proposition2:cost}
	With known $\btheta$, the optimization problem in Eq.~\eqref{general:proposition1} could be solved with a convex-concave procedure (CCCP):
	$$\min_{\bw^l, \bgamma^l} \mathcal{L}(\bw^l, \bgamma^l)$$
	with the cost function:
	with 
	{
		\fontsize{8.5}{2}
	\begin{equation}
    \begin{aligned}
    \mathcal{L}(\bw^l, \bgamma^l) \define & {\bw^l}^{\top} \Hessian^l \bw^l+ 2 {\bw^l}^{\top} \left[\Gradient^l-\Hessian^l {\bw^l}^{*}\right]  \\
		&+ {\bw^l}^\top {\bG^l}^{-1} \bw^l  + \log |\bG^l| + \log |\Hessian^l+ {\bG^l}^{-1} |
    \label{eq:general:cost:cccp}
    \end{aligned}
    \end{equation}}
\end{prop}
% The detailed proof for proposition~\ref{general:proposition1:cost} and  proposition~\ref{general:proposition2:cost} is in Section~\ref{app:proof for proposition 1} and Section~\ref{app:proof for proposition 2} of Appendix. 
With proposition~\ref{general:proposition2:cost}, we also give the analytic form for $\bgamma^l$ and $\bw^l$ based on the chain rule and basic principle for convex analysis. We give the iterative analytic form directly as Eq.~\eqref{eq:cccp_1}-Eq.~\eqref{eq:cccp_4}. 
% The detailed derivation can be found in Appendix~\ref{appsec:cccp_procedure}.
\begin{figure}[t]
	\begin{center}
		\includegraphics[scale = 0.24]{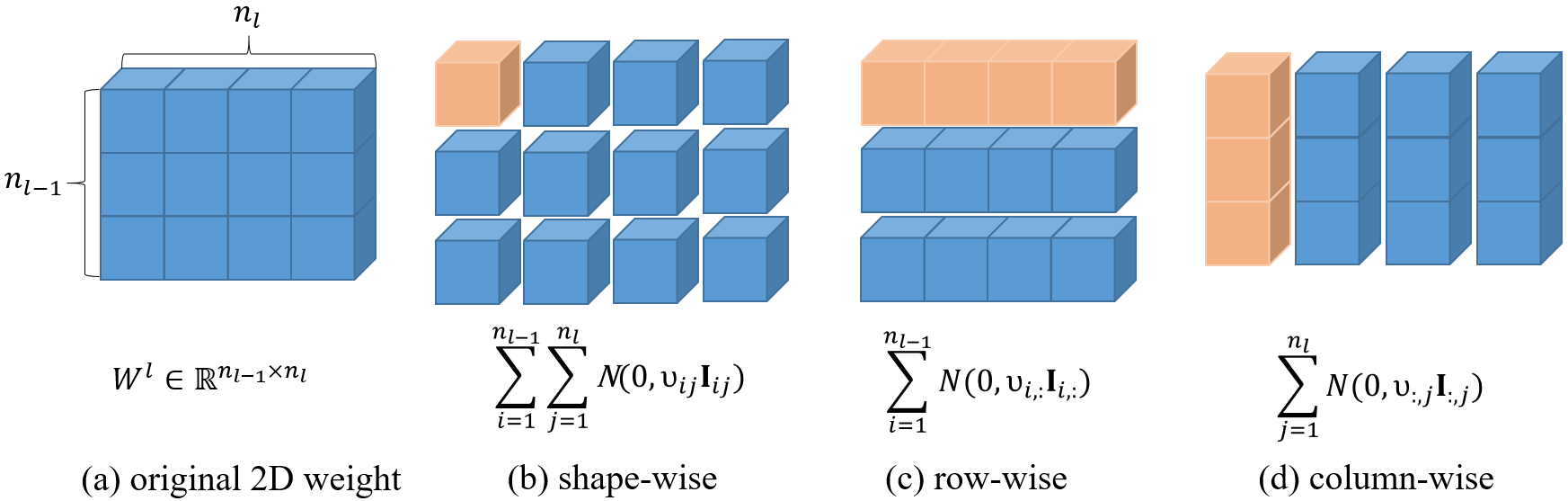}
		\caption{Structural sparsity for 2-D weight matrices} 
		\label{fig:mlp_shape}
	\end{center}
\end{figure}
\begin{align}
&C^l(t)  = \left({\bG^l(t)}^{-1}+{\Hessian}({\bw^l}^*, \btheta^*)\right)^{-1} 
\label{eq:cccp_1}
\\
&{\alpha_{ij}^l}(t) = { -\frac{\mathbf{C}^l_{ij}(t)}{\br^l_{ij}(t)^2} +\frac{1}{\br^l_{ij}(t)}}
\label{eq:cccp_2}
\\
&\br^l_{ij}(t+1) = \frac{|\bw^l_{ij}(t+1)|}{\omega^l_{ij}(t)}
\label{eq:cccp_3}
\end{align}
where $C$ and $\alpha(t)$ are the introduced intermediate variables; $\omega_{ij}$ is assumed as \begin{math}
\omega_{ij} \define \sqrt{\alpha_{ij}}
\end{math}. For $\bw^l$, it could be updated according to:
{
	%	\small
	\fontsize{9.0}{2}
	\begin{equation}
	\begin{aligned}
	\bw^l(t+1)  = &\argmin\limits_{\bw^l}
	{\bw^l}^{\top} \Hessian^l \bw^l+ 2{\bw^l}^{\top}\left(\Gradient^l-\Hessian^l {\bw^l}^{*}\right) \\ &+2\sum_{i=1}^{n_{l-1}}\sum_{j=1}^{n_{l}} \|\omega^l_{ij}(t) \cdot \bw^l_{ij}\|_{\ell_1}\\
	\approx&\argmin\limits_{\bw^l} E(\cdot) + R(\omega^l(t) \circ {\bw}^l(t))
% 	\approx&\argmin\limits_{\bw^l} E(\cdot) + 2\sum_{i=1}^{n_{l-1}}\sum_{j=1}^{n_{l}} \|\omega^l_{ij}(t) \cdot \bw^l_{ij}\|_{\ell_1}
	\label{eq:cccp_4}
	\end{aligned}
	\end{equation}
} where $R(\cdot)$ stands for the regularization item.

\subsection{Algorithm}
By introducing the Bayesian framework, we propose a sparse Bayesian deep learning algorithm. The pseudo-code is summarized in Algorithm~\ref{algo:algorithm}.

\begin{algorithm}[t] 
	\caption{Sparse Bayesian Deep learning algorithm}
	\label{algo:algorithm}
	\textbf{Initialize:} hyper-parameters ${\omega}^l(0), {\Upsilon}^l(0) = 1$; regularization tuning parameter $\lambda^l \in \R^{+}$; threshold for NN pruning $\kappa_\upsilon, \kappa_w \in \R^{+}$; where $l = 1,\ldots,L$;
	
	\textbf{for} {$t=1$ to $T_{\max}$}
	\begin{enumerate} 
		\item Maximum likelihood with regularization:
		\begin{align}
		\label{algo3:eq1}
		\min\limits_{{\bw}} {E}(\cdot) +  \sum\limits_{l=1}^{L} \lambda^l R(\omega^l(t) \circ {\bw}^l(t))
		\end{align}
		\item Update hyper-parameters $\bGamma(t)^l, \alpha(t)^l, \omega(t)^l, \forall l = 1,\ldots,L$, as Table~\ref{tab:mlp}.
		\item Dynamic pruning:
		
		\textbf{if} $\upsilon_{ij}(t)<\kappa_{\upsilon}$ or $|\bw_{ij}(t)|<\kappa_{w}$ \textbf{then}	
			
			\qquad \textbf{prune} $\bw_{ij}(t)$		
			
		\textbf{end if}
	\end{enumerate} 
	\textbf{end for} 
\end{algorithm}
The algorithm is consisting of three parts.
The first part is a reweighted optimization, where $E(\cdot)$ is usually known as the loss function in a neural network as mean square error (MSE) and root mean square error (RMSE). $R(\cdot)$ stands for the regularization item. Its analytical form is in Table. $\lambda$ is a tradeoff parameter which denotes the regularization tuning parameter for $R(\cdot)$. And $\bw$ will be updated in this part. 
The second part is to update the hyper-parameters, including $\Upsilon$, $\alpha$ and $\omega$. They will be updated each iteration with $T_{max}$ being the index of maximal iteration. 
The detailed update strategy can be found in Table~\ref{tab:mlp}.
The detailed update strategy can be found in Appendix~\ref{appsec:cccp_procedure} and Table~\ref{tab:mlp}.
The third part is for pruning. Unimportant neurons will be identified and removed from the original neural network. In this paper, we propose a dynamic pruning strategy which considers both the uncertainty and the magnitude of the identified parameter. Two threshold values $\kappa_{\upsilon}$ and $\kappa_{w}$ are defined in the beginning. The connections with $\upsilon_{ij}$ smaller than $\kappa_{\upsilon}$ or $|\bw_{ij}|$ smaller than $\kappa_{w}$ will be removed in each iteration. 
% Above procedures will be implemented iteratively until satisfying some stopping criterion. 
Besides, in order to promote the structured sparsity~\citep{wenwei2016} for neural network, we also propose some examples of structured sparsity as shown in Fig.~\ref{fig:mlp_shape}.
The corresponding regularization items and hyperparameter update method are shown in Table~\ref{tab:mlp}(b)-(c). 

\begin{table*}
	\caption{Hyper-parameter update rule in Algorithm\ref{algo:algorithm}}
	\label{tab:mlp}
	\begin{center}
		 \resizebox{1.00\textwidth}{!}{
		\begin{tabular}{|c|c|c|c|c|}
			\hline
			Category  & Sparse prior     & $R^{\layerindex}(\omega\circ\bw)$  & ${\omega}^{\layerindex}$    & ${\Upsilon}^{\layerindex}$  \\ \hline
			(a) Shape-wise & $\prod\limits_{i}\prod\limits_{j}\mathcal{N}(\mathbf{0}, \Upsilon_{i,j}\mathbf{I}_{i,j}) $               
			& $\sum\limits_{i = 1}^{i}\sum\limits_{j = 1}^{j}\|\omega_{{i},{j}}^{\layerindex}(t) \circ \bw_{{i},{j}}^{\layerindex}(t)\|_{\normone}$                    
			& \begin{tabular}[c]{@{}c@{}}
				${\omega}^{\layerindex}_o(t) = \sqrt{\sum\limits_{i}\sum\limits_{j}|\alpha_{{i},{j}}^{\layerindex}(t)|}$\\ 
				${\omega}_{{i},{j}}^{\layerindex}(t) = {\omega}^{\layerindex}_o(t) \cdot \rmI_{{i},{j}}^{\layerindex}$
			\end{tabular}                                                                       
			& \begin{tabular}[c]{@{}c@{}}
				${\Upsilon}^{\layerindex}_o(t) = \frac{\|\bw_{{i},{j}}^{\layerindex}(t)\|_2}{\omega_{{i},{j}}^{\layerindex}(t-1)}$\\ 
				$\Upsilon_{{i},{j}}^{\layerindex}(t) = {\Upsilon}^{\layerindex}_o(t) \cdot \rmI_{{i},{j}}^{\layerindex}$
			\end{tabular}  \\ \hline
			(b) Row-wise  & $\prod\limits_{i}\mathcal{N}(\mathbf{0}, \Upsilon_{i,:}\mathbf{I}_{i,:}) $ 
			& $\sum\limits_{i = 1}^{i}\|\omega_{{i},:}(t)\circ \bw_{{i},:}^{\layerindex}(t)\|_{\normone}$       
			& \begin{tabular}[c]{@{}c@{}}
				${\omega}^{\layerindex}_o(t) = \sqrt{\sum\limits_{i}{|\alpha_{{i},:}^{\layerindex}(t)|}}$\\ 
				${\omega}_{{i},:}^{\layerindex}(t) = {\omega}^{\layerindex}_o(t) \cdot \rmI_{ {i},:}^{\layerindex}
				$\end{tabular}                                                                        
			& \begin{tabular}[c]{@{}c@{}}
				${\Upsilon}^{\layerindex}_o(t) = \frac{\|\bw_{{i},:}^{\layerindex}(t)\|_2}{\omega_{{i},:}^{\layerindex}(t-1)}$\\ 
				$\Upsilon_{{i},:}^{\layerindex} = {\Upsilon}^{\layerindex}_o(t) \cdot \rmI_{{i},:}^{\layerindex}$
			\end{tabular}                                                                  
			\\ \hline
			(c) Column-wise &    $\prod\limits_{j}\mathcal{N}(\mathbf{0}, \Upsilon_{:,j}\mathbf{I}_{:,j}) $            
			& $\sum\limits_{j = 1}^{j}\|\omega_{:,{j}}^{\layerindex}(t) \circ \bw_{:,{j}}^{\layerindex}(t)\|_{\normone}$                                                                     
			& \begin{tabular}[c]{@{}c@{}}
				${\omega}^{\layerindex}_o(t) = \sqrt{\sum\limits_{j}{|\alpha_{:,{j}}^{\layerindex}(t)|}}$\\ 
				${\omega}_{:,{j}}^{\layerindex}(t) = {\omega}^{\layerindex}_o(t) \cdot \rmI_{:,{j}}^{\layerindex}$
			\end{tabular}                                                                              
			& \begin{tabular}[c]{@{}c@{}}
				${\Upsilon}^{\layerindex}_o(t) = \frac{\|\bw_{:,{j}}^{\layerindex}(t)\|_{2}}{\omega_{:,{j}}^{\layerindex}(t-1)}$\\ 
				$\Upsilon_{:,{j}}^{\layerindex}(t) = {\Upsilon}^{\layerindex}_o(t) \cdot \rmI_{:,{j}}^{\layerindex}$
			\end{tabular}  \\ \hline
		\end{tabular}
	}
	\end{center}
\end{table*}

\section{Experiment}
\label{sec:experiment}
We design FC networks with diverse settings for the number of hidden layers and hidden neurons for different benchmarks. The dataset for training is also selected with different ratios of original dataset.
In this section, one-step ahead prediction experiment and simulation experiment were implemented on Cascaded Tanks System (2016)\citep{cascad_benchmark}. The high-quality and well-described datasets are provided in \url{http://www.nonlinearbenchmark.org/index.html}.

\subsubsection{Benchmark description}
The cascaded tanks system is a liquid level control system. 
Its mathematical model can be constructed as:
\begin{equation}
\begin{aligned}
\label{eq:cascad_tank} 
& \dot{x_1}(t) = -k_1\sqrt{{x_1}(t)} + k_4 u(t) + w_1(t),
\\
& \dot{x_2}(t) = k_2\sqrt{{x_1}(t)} - k_3\sqrt{{x_2}(t)} + w_2(t),
\\
& y(t) = {x_2}(t) + e(t)
\end{aligned}
\end{equation}
where $u(t)$ is the input pump voltage, 
% signal
$y(t)$ is the output which measures the liquid level, $x_1(t)$ and $x_2(t)$ are the states of the system, $w_1(t)$, $w_2(t)$ and $e(t)$ are the noise and $k_1,k_2,k_3$, and $k_4$ are the constants which is decided by the system properties. 
In the benchmark provided by \url{http://www.nonlinearbenchmark.org/index.html}, both training data and test data include 1024 samples. The root of mean square error between predicted output $\hat{y}$ and true output $y$ is selected as the evaluation criterion:
\begin{equation} 
E(\cdot) \define RMSE(\hat{y},y) = \sqrt{\frac{1}{1024}\sum\limits_{i=1}^{1024} ({y_i - \hat{y}_i})^2}
\label{eq:rmse_cad}
\end{equation}
\subsubsection{One-step ahead prediction}
A Fully Connected network is initialized to be with two hidden layers and $100$ neurons in each layer. The data lags for both input and output are set as $5$. We apply row-wise (Fig.~\ref{fig:mlp_shape}(c)) and column-wise (Fig.~\ref{fig:mlp_shape}(d)) regularization with different $\lambda$ to the weight matrices. Experiments are implemented with different ratios of the original training dataset. The ratio is selected in the scope $[5\%, 10\%, 20\%, 30\%, 40\%, 50\%, 60\%, 70\%, 80\%, 90\%, 100\%]$. As a comparison, we also experimented without applying regularization with the same network structure. Each experiment with a different setting was repeatedly implemented $50$ times in total. 

\begin{figure}[htbp]
\centering    
\subfigure[best prediction error]{
 \label{fig:cad_best_preciction}     
\includegraphics[scale=0.22]{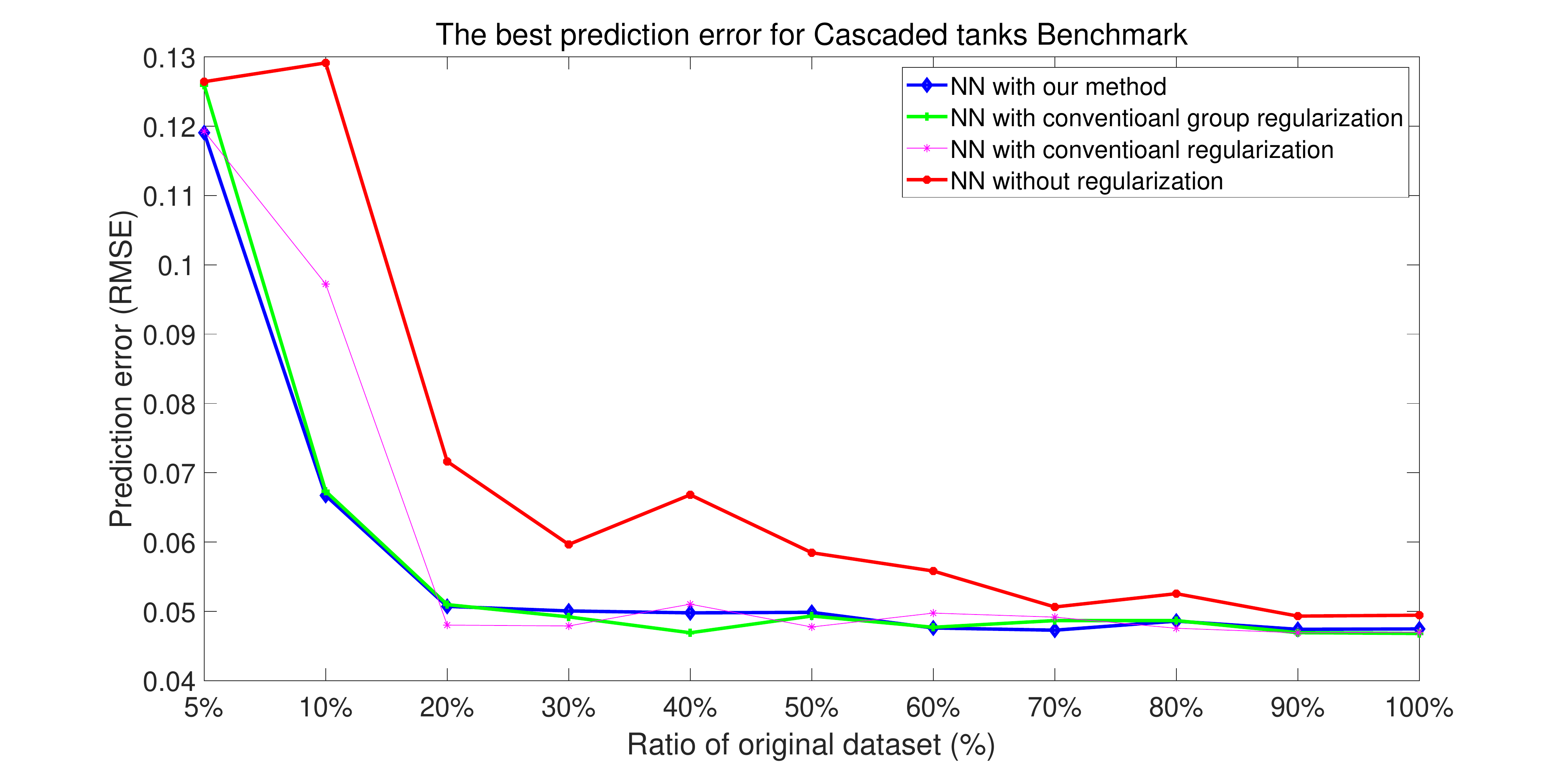}
}   
\subfigure[mean prediction error]{ 
\label{fig:cad_mean_preciction}     
\includegraphics[scale=0.22]{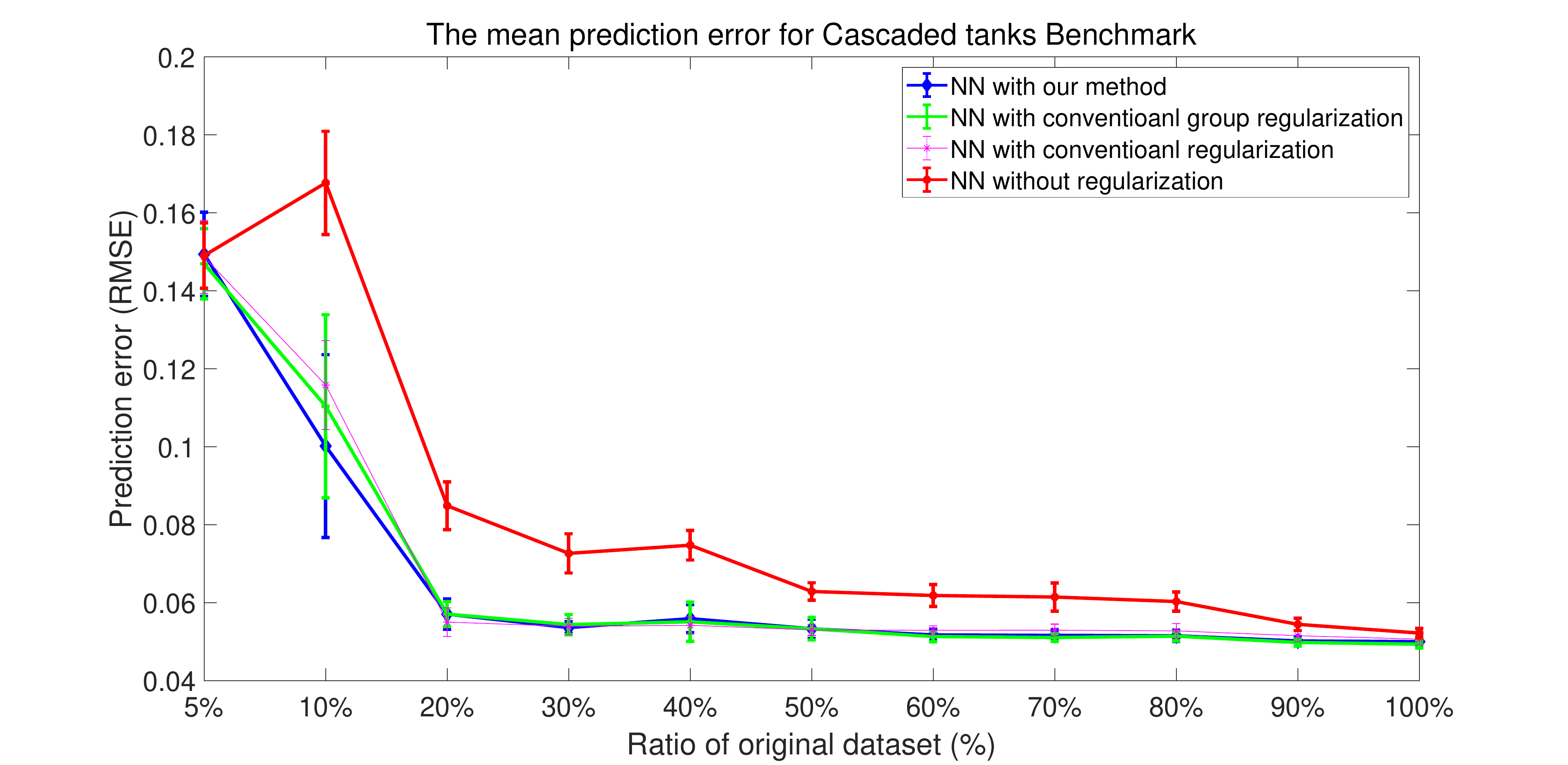}
}    
\caption{The predicted result with different ration of original dataset for cascaded tanks system. 
The comparison of four approaches is shown (i.e. our method with Bayesian method, neural network with conventional group regularization, neural network with conventional regularization, only neural network without regularization)
Subfigure.(a) shows the best prediction error and Subfigure.(b) shows the mean prediction error with vertical error bar at each data point. In both subfigures, the red curve represents the result of our method, the green curve is the result of conventional group regularization method and and blue curve stands for the result of neural network without regularization. 
$11$ data points in each curve correspond to the best prediction error with $11$ diverse ratios from $5\%$ to $100\%$.} 
\label{fig:prediction_result_small_data}
\end{figure}

The result is shown in Fig.~\ref{fig:prediction_result_small_data}, where the smallest and mean prediction errors with a different dataset of four approaches are plotted in different colours (i.e. our method with Bayesian method, neural network with conventional group regularization, neural network with conventional regularization, only neural network without regularization). From the result, we observe some interesting phenomenon. First of all, all curves show a similar and reasonable trend that the prediction error became smaller with more provided training data.
Secondly, it is obvious that compared to the experiment without regularization, our method could obtain a smaller prediction error with different ratio of original dataset. In the experiments with conventional group regularization, we also applied row-wise (Fig.~\ref{fig:mlp_shape}(c)) and column-wise (Fig.~\ref{fig:mlp_shape}(d)) regularization with different $\lambda$ to the weight matrices. Our method could also achieve comparable result in terms of best and mean prediction error. The result shows the proposed method has the capacity to approximate the cascaded tanks system. 

Besides, as shown in the blue curve, the prediction error shows a convergence trend. And the optimal prediction error is $0.0472$ with only $70\%$ dataset. We argue that there exits a balance between the number of samples and model accuracy.
On one hand, if the provided data is too less (e.g.less than $50\%$), we cannot obtain an optimal model no matter whether we introduce the regularization. On the other hand, the result shows that it is possible to obtain the optimal model with a reduced training dataset. 
On the whole, this result shows that our method can keep good performance even without enough data. 
Last but not least, the sparsity of weight matrices also changes over iterations. 
And compared with conventional group regularization method, our method could achieve a relative sparser model especially when the training data is more than $60\%$. For example, with $80\%$ training data, the sparsity of the model with our method and conventional group regularization method is $2.5\%$ and $10.98\%$. A more direct comparison of the model sparsity for the proposed method and conventional group regularization method, refer to Table.~\ref{tab:comparision_cad_sparsity}. 
\begin{table*}[ht]
\caption{Comparison of model sparsity on Cascaded tanks Benchmark}
\label{tab:comparision_cad_sparsity}
\begin{tabular}{|c|c|c|c|c|c|c|c|c|c|c|c|}
\hline
Ratio                                                                                  & 5\%     & 10\%   & 20\%   & 30\%   & 40\%   & 50\%   & 60\%   & 70\%   & 80\%    & 90\%    & 100\%  \\ \hline
\begin{tabular}[c]{@{}c@{}}NN with conventioanl \\group regularization\end{tabular} & 46.36\% & 3.27\% & 2.23\% & 2.78\% & 1.94\% & 3.04\% & 3.91\% & 3.18\% & 10.98\% & 51.26\% & 9.78\% \\ \hline
\begin{tabular}[c]{@{}c@{}}Our method with \\ Bayesian framework\end{tabular}               & 61.52\% & 2.01\% & 2.95\% & 3.42\% & 2.14\% & 2.71\% & 2.79\% & 2.30\% & 2.50\%  & 9.79\%  & 2.23\% \\ \hline
\end{tabular}
\end{table*}

\subsubsection{Simulation prediction}
\begin{figure}[ht]
	\centering
	\includegraphics[scale=0.23]{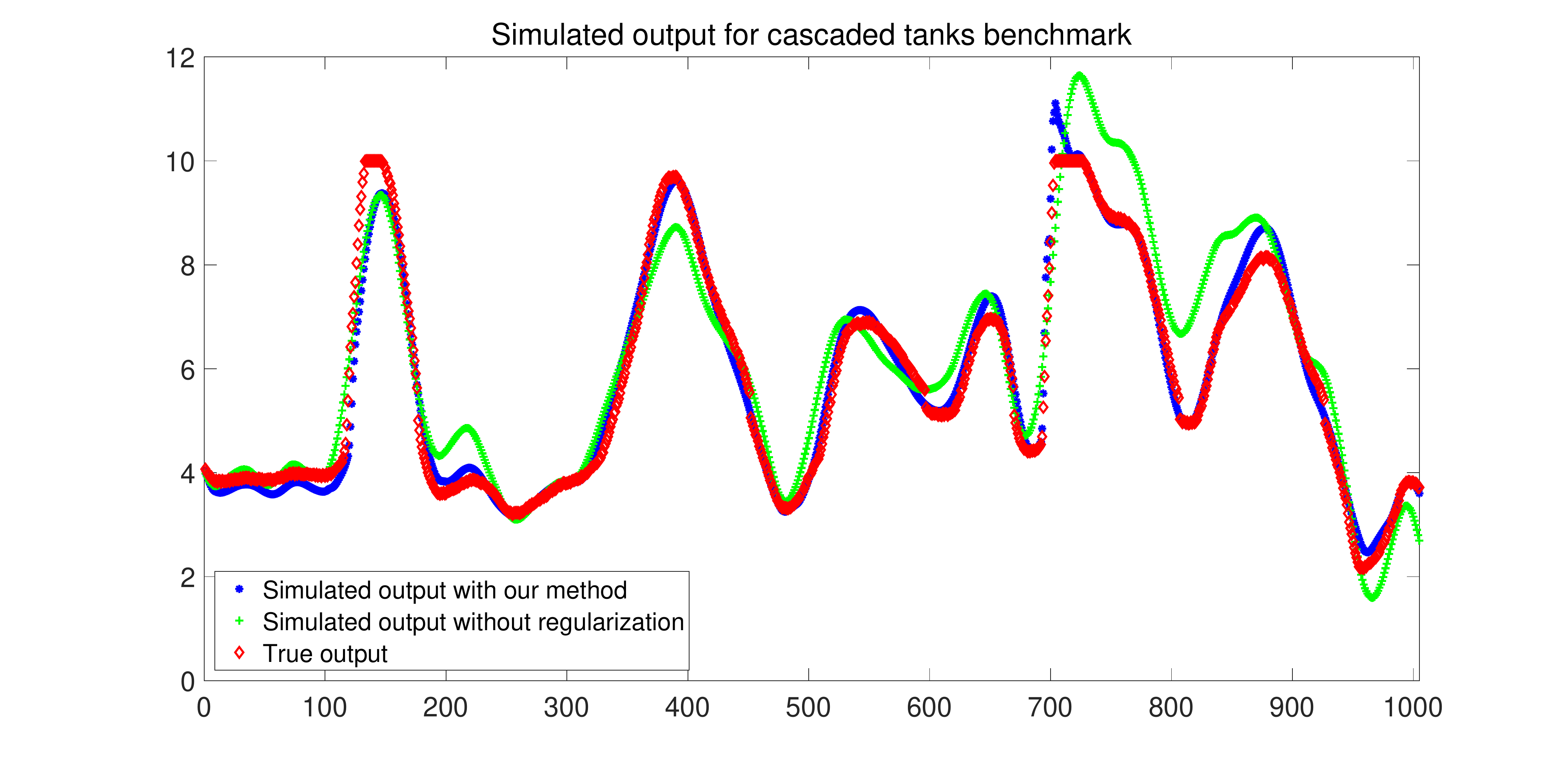}
	\caption{A snapshot of simulated output for cascaded tanks system.}
	\label{fig:cad_simulation_output}
\end{figure}

Now we consider the more challenging task for simulation. With the same evaluation criterion as Eq.~\eqref{eq:rmse_cad}, a FC network with three hidden layers and $10$ neurons in each layer is initialized. The data lags for both input and output are set as $19$. We apply shape-wise (Fig.~\ref{fig:mlp_shape}(b)), row-wise (Fig.~\ref{fig:mlp_shape}(c)), column-wise (Fig.~\ref{fig:mlp_shape}(d)) regularization with different $\lambda$ to the weight matrices. With the experiment setting, the best simulation error of our method is $0.344$. 
The output is shown in Fig.~\ref{fig:cad_simulation_output}, where the bold blue and red curves stand for the simulation output and ground truth respectively. 
For comparison, we also did the experiment without applying regularization on weight matrices with the same network structure. This experiment was repeatedly implemented with different initialization for $20$ times in total. Finally, the simulation errors with mean value as $2.0634$ and standard deviation as $1.6375$ for these $20$ repeated experiments were received. We showed ten of them in Fig.~\ref{fig:cad_simulation_output} and the output with the smallest simulation error ($0.777$) was plotted with bold green line. It is obvious that our proposed method has better capacity to approximate the cascaded tanks system. 
We also make a comparison with other approaches as shown in Table.~\ref{tab:comparision_cad_simulation}, which shows that our method could achieve a comparable result.

\begin{table}[ht]
	\centering
	\caption{Comparison on simulation error on Cascaded tanks Benchmark}
	\begin{tabular}{ll}
		\hline
		Method                                                                  & RMSE  \\ \hline
		NOMAD algorithm~\citep{oatao17993}                                                                 & 0.376 \\
		PWARX model~\citep{MATTSSON201840}                                                             & 0.350 \\ 
		
		Volt.FB~\citep{schoukens2016modeling}                                                                 & 0.397 \\
		Poly.FB~\citep{schoukens2016modeling}                                                                  & 0.487 \\
		LTI~\citep{schoukens2016modeling}                                                                      & 0.588 \\
		\hline
		DNN without regularization                                                             & \textbf{0.777} \\ %\hline
		Our Method with regularization                                                             & \textbf{0.344} \\ \hline
	\end{tabular}
\label{tab:comparision_cad_simulation}
\end{table}

\section{Conclusion}
\label{sec:conclusion}
This paper proposes a sparse Bayesian deep learning approach for nonlinear system identification.
By promoting sparsity in the neural network model, the proposed method could overcome the overfitting problem.
It was also proved that this method could achieve similar prediction accuracy to standard identification techniques with a small dataset.
For the simulation, we also could achieve competitive result, which supports that our method could approximate the intrinsic feature of a system. 
%\begin{ack}
%The work of Hongpeng Zhou is sponsored by the program of China Scholarships Council (No.201706120017).
%\end{ack}
%\bibliography{ifacconf}             % bib file to produce the bibliography% with bibtex (preferred)

%\bibliography{ref}  

\appendix
\newpage
\onecolumn
\section{Proof for proposition~\ref{general:proposition1:cost}}
\label{app:proof for proposition 1}
% \wei{you should not use $E$, the general likelihood. You should use Gaussian. This is complicated, but you have to do that layer b layer and try to derive yourself, not just copy my thesis, right? So in short, as a special case, change $E$ to Gaussian, and derive yourself please, then the Hessian everything is analytical}
% \wei{and I didn't see anything related to layer, where is $l$. you do that layer by layer, right? but this is missing, it seems that you do everything in one go, you didn't use backpropgation, you should show that. Please don't copy my latex in my thesis. If you can't derive everything, you will have trouble in the future}

\begin{pf}	
Given the likelihood with Gaussian distribution $$p(\by|{\bw}^l,\sigma^2) =  (2\pi\sigma^2)^{\frac{N}{2}}\exp\left(-\frac{1}{2\sigma^2}{\sum_{t=1}^{N}\left(y(t)-\text{Net}\left(\bz(t),{\net}\right)\right)^2}\right)$$ as in~\eqref{general-likelihood-expo} and sparse prior with super Gaussian distribution $$p({\bw}^l, {\bG^l}) =\prod_{i=1}^{n_{l-1}}\prod_{j=1}^{n_l}\bN({\bw^l}_{ij}|0,\br_{ij}^l){p}({\hyper_{ij}^l}) = \bN({\bw}^l|\mathbf{0},{\bG^l}){p}({\bG^l})$$ as in~\eqref{eq:super gaussian prior}, we go straightly into the marginal likelihood:
\begin{equation}
\begin{aligned}
&p(\by) = \int p(\by|{\bw}^l)p({\bw}^l)d{\bw}^l \\
= & \int p({\by}|{\bw^l}, \btheta) \bN( {\bw^l}|\mathbf{0},{\bG^l})\prod_{i=1}^{n_{l-1}}\prod_{j=1}^{n_{l}}\prior({\hyper_{ij}^l})d{\bw}^l \\
= & {(2\pi\sigma^2)^{\frac{N}{2}}} \int \exp\{-\energy\}\bN( {\bw^l}|\mathbf{0},{\bG^l})\prod_{i=1}^{n_{l-1}}\prod_{j=1}^{n_{l}}\prior({\hyper_{ij}^l})d{\bw}^l
\label{general:generalmarignal}
\end{aligned}
\end{equation}
To compute the intractable integral in Eq.~\eqref{general:generalmarignal}, we adopt the Laplace approximation method. For easy notation, we use $E({\bw},\sigma^2)$ to denote \begin{math}
	\energy
\end{math}. It is worth noting that $E(\cdot)$ is usually known as the loss function in a neural network. Suppose $\btheta$ is known and $E({\bw},\sigma^2)$ could be expanded around some point ${\bw^l}^{*}$ according to Talylor series expansion:
\begin{equation}
\begin{aligned}
E({\bw^l},\sigma^2)
\approx E({\bw^l}^*, \btheta)+({\bw^l}-{\bw^l}^*)^{\top}\mathbf{G}({\bw}^*, \btheta)+\frac{1}{2}({\bw^l}-{\bw^l}^*)^{\top} \mathbf{H}({\bw}^*, \btheta) ({\bw^l}-{\bw^l}^*) 
\label{eq:taylor_energy_function}
\end{aligned}
\end{equation}
where $\mathbf{G}(\cdot)$ is the gradient and $\mathbf{H}(\cdot)$ is the Hessian of $E(\cdot)$ to ${{\bw^l}^{*}}$. Specifically, according to the principle of backward propagation and chain rule, the iterative analytical form of $\Gradient(\cdot)$ for a specific weight matrix $\bw^l$, could be given as follows:
\begin{equation}
\begin{aligned}
\Gradient({\bw^l}) &= \frac{\partial E(\cdot)}{\partial {\bw^l}} = \frac{\partial E(\cdot)}{\partial a^l}\circ\frac{\partial a^l}{\partial h^l}\frac{\partial h^l}{\partial {\bw^l}} = \frac{\partial E(\cdot)}{\partial a^l}\circ{f'(h^l)}{a^{l-1}}^\top \\
\frac{\partial E(\cdot)}{\partial a^l} &= \frac{\partial E(\cdot)}{\partial a^{l+1}}\circ\frac{\partial a^{l+1}}{\partial h^{l+1}}\frac{\partial h^l}{\partial {\bw}^{l+1}} = \frac{\partial E(\cdot)}{\partial a^{l+1}}\circ{f'(h^{l+1})}{{\bw}^{l+1}}^\top
\end{aligned}
\label{eq:fc_gradient}
\end{equation}
For Hessian, a recursive method is proposed in \cite{botev2017practical} to compute $\Hessian(\cdot)$ for a FC layer:
\begin{equation}
\begin{aligned}
\Hessian({\bw^l}) =& a^{l-1} \cdot({a^{l-1}})^\top \otimes H^l, \ \ \ 
H^l = B^l {W^{l+1}}^T H^{l+1} W^{l+1} B^l + D^l,  \\
B^l =& \diag(f'(h^l)), \ \ \
D^l = \diag(f''(h^l) \frac{{\partial}E(\cdot)}{{\partial}a^l})
\end{aligned}
\label{eq:fc_hessian} 
\end{equation}
where $\otimes$ stands for Kronecker product; $H^l$ denotes the pre-activation Hessian which needs to be computed recursively for each layer.
To derive the cost function in~\eqref{general:proposition1}, we introduce the posterior mean and covariance
\begin{subequations}
	\begin{align}
	\mean_{{\bw^l}}&= {{\Sigma}_{{\bw^l}}} \cdot \left[\Gradient({\bw^l}^*, \btheta) + \Hessian({\bw^l}^*, \btheta) {\bw^l}^{*} \right], \label{general:prior:mean} \\
	{{\Sigma}_{{\bw^l}}}&= \left[\Hessian({\bw^l}^*, \btheta)+ {\bG^l}^{-1} \right]^{-1}. \label{general:prior:variance} 
	\end{align}
	\label{app:general:prior:moment} 
\end{subequations}
According to Eq.~\eqref{eq:taylor_energy_function}, the likelihood $p({\by}|{\bw^l}, \btheta)$ could be approximated as:
\begin{equation}
\begin{aligned}
& p({\by}|{\bw^l}, \btheta) \\
=& {(2\pi\sigma^2)^{\frac{N}{2}}} \cdot \exp\{-\energy\} \\
\approx & {(2\pi\sigma^2)^{\frac{N}{2}}} \cdot \exp \left \{-\left(\quadratic+E({\bw}^*, \btheta)\right)\right \} \\
= & {(2\pi\sigma^2)^{\frac{N}{2}}} \cdot \exp \left \{-\left(\frac{1}{2}{\bw^l}^{\top} \Hessian({\bw^l}^*, \btheta) {\bw^l}+ {\bw^l}^{\top} \left[\Gradient({\bw^l}^*, \btheta)-\Hessian({\bw^l}^*, \btheta) {\bw^l}^{*}\right]\right) \right \} \\
& \cdot \exp \left \{- \left(\frac{1}{2}{{\bw^l}^*}^{\top} \Hessian({\bw^l}^*, \btheta) {{\bw^l}^*} - {{\bw^l}^*}^{\top} \Gradient({\bw^l}^*, \btheta)  + E({\bw}^*, \btheta) \right) \right \} \\
= & {(2\pi\sigma^2)^{\frac{N}{2}}} \cdot b({\bw^l}^*, \btheta) \cdot \exp \left \{-\left(\frac{1}{2}{\bw^l}^{\top} \Hessian({\bw^l}^*, \btheta) {\bw^l}+ {\bw^l}^{\top} \hat{\Gradient}({\bw^l}^*, \btheta)\right) \right \} 
\label{likelihood-approximation}
\end{aligned}
\end{equation}
where
\begin{equation}
\begin{aligned}
b({\bw^l}^*, \btheta) & \define \exp \left \{- \left(\frac{1}{2}{{\bw^l}^*}^{\top} \Hessian({\bw^l}^*, \btheta) {{\bw^l}^*} - {{\bw^l}^*}^{\top} \Gradient({\bw^l}^*, \btheta)  + E({\bw}^*, \btheta) \right) \right \} 
\\
\hat{\Gradient}({\bw^l}^*, \btheta)& \define \Gradient({\bw^l}^*, \btheta)-\Hessian({\bw^l}^*, \btheta) {\bw^l}^{*}.
\notag
\end{aligned}
\end{equation}

We can write the approximate marginal likelihood as
\begin{equation}
\begin{aligned}
& \int p({\by}|{\bw^l}, \btheta) \bN( {\bw^l}|\mathbf{0},{\bG^l})\prod_{i=1}^{n_{l-1}}\prod_{j=1}^{n_{l}}\prior({\hyper_{ij}^l})d{\bw^l} \\
= & {(2\pi\sigma^2)^{\frac{N}{2}}} \cdot b({\bw^l}^*, \btheta) \cdot \int \exp \left \{-\left(\frac{1}{2}{\bw^l}^{\top} \Hessian({\bw^l}^*, \btheta) {\bw^l}+ {\bw^l}^{\top} \hat{\Gradient}({\bw^l}^*, \btheta)\right) \right \}\bN( {\bw^l}|\mathbf{0},{\bG^l})\prod_{i=1}^{n_{l-1}}\prod_{j=1}^{n_{l}}\prior({\hyper_{ij}^l})d{\bw^l} \\
=&\frac{{(2\pi\sigma^2)^{\frac{N}{2}}} \cdot b({\bw^l}^*, \btheta)}{\left(2\pi\right)^{(n_l\times n_{l-1})/2}|{\bG^l}|^{1/2}}
\int \exp\{-\hat{E}({\bw^l}, \btheta)\}d{\bw^l} \prod_{i=1}^{n_{l-1}}\prod_{j=1}^{n_{l}}\prior({\hyper_{ij}^l})
,
\label{general:prior:integral}
\end{aligned}
\end{equation}
where
\begin{equation}
\begin{aligned}
\hat{E}({\bw^l})=\frac{1}{2}{\bw^l}^{\top} \Hessian({\bw^l}^*, \btheta) {\bw^l}+ {\bw^l}^{\top} \hat{\Gradient}({\bw^l}^*, \btheta)+\frac{1}{2} {\bw^l}^\top {\bG^l}^{-1} {\bw^l}.
\end{aligned}
\end{equation}
Equivalently, we get
\begin{equation}
\begin{aligned}
\hat{E}({\bw^l})=\frac{1}{2}({\bw^l}-\mean_{\bw^l})^\top{{\Sigma}_{{\bw^l}}}^{-1}({\bw^l}-\mean_{\bw^l})+\hat{E}({\mathbf{y}}),
\label{general:prior:integral2}
\end{aligned}
\end{equation}
where the data-dependent term can be re-expressed as
\begin{equation}
\begin{aligned}
\hat{E}({\by})
=&  {\frac{1}{2}\mean^{\top} \Hessian({\bw^l}^*, \btheta) \mean+  \mean^{\top}\Gradient({\bw^l}^*, \btheta)}+ \frac{1}{2}\mean^{\top}{\bG^l}^{-1}\mean \\
=&\min_{{\bw^l}}  \left[\frac{1}{2}{\bw^l}^{\top} \Hessian({\bw^l}^*, \btheta) {\bw^l}+ {\bw^l}^{\top} \hat{\Gradient}({\bw^l}^*, \btheta) + \frac{1}{2}{\bw^l}^\top {\bG^l}^{-1} {\bw^l}\right] \\
=&\min_{{\bw^l}}  \left[\frac{1}{2}{\bw^l}^{\top} \Hessian({\bw^l}^*, \btheta) {\bw^l}+ {\bw^l}^{\top} \left(\Gradient({\bw^l}^*, \btheta)-\Hessian({\bw^l}^*, \btheta) {\bw^l}^{*}\right) + \frac{1}{2}{\bw^l}^\top {\bG^l}^{-1} {\bw^l}\right].
\label{general:prior:data-dependent-term}
\end{aligned}
\end{equation}

Using~\eqref{general:prior:integral2}, we can evaluate the integral in~\eqref{general:prior:integral} to obtain
\begin{equation}
\begin{aligned}
\int \exp \left \{-\hat{E}({\bw^l}) \right \}d{\bw^l}=\exp \left \{-\hat{E}({\mathbf{y}}) \right \} (2\pi)^{{n_l}\times{n_{l-1}}}|{{\Sigma}_{{\bw^l}}}|^{1/2}.
\end{aligned}
\end{equation}
Applying a $-2\log(\cdot)$ transformation to~\eqref{general:prior:integral}, we have
\begin{equation}
\begin{aligned}
&-2\log\left[\frac{{(2\pi\sigma^2)^{\frac{N}{2}}} \cdot b({\bw^l}^*, \btheta)}{\left(2\pi\right)^{{n_l}\times{n_{l-1}}/2}|{\bG^l}|^{1/2}}
\int \exp\{-\hat{E}({\bw^l})\}d{\bw^l} \prod_{i=1}^{n_{l-1}}\prod_{j=1}^{n_l}\prior(\hyper_{ij}^l)\right] \\
\propto & -2\log {{(2\pi\sigma^2)^{\frac{N}{2}}} \cdot b({\bw^l}^*, \btheta)} + \hat{E}({\mathbf{y}}) +\log  |{\bG^l}|+\log |\Hessian({\bw^l}^*, \btheta)+ {\bG^l}^{-1} | - 2 \sum_{i=1}^{n_{l-1}}\sum_{j=1}^{n_l} \log \prior(\hyper_{ij}^l)\\
\propto & {\bw^l}^{\top} \Hessian({\bw^l}^*, \btheta) {\bw^l}+ 2{\bw^l}^{\top} \hat{\Gradient}({\bw^l}^*, \btheta)+ {\bw^l}^\top {\bG^l}^{-1} {\bw^l} +\log  |{\bG^l}|+\log |\Hessian({\bw^l}^*, \btheta)+ {\bG^l}^{-1} | \\
& -2\log {{(2\pi\sigma^2)^{\frac{N}{2}}} \cdot b({\bw^l}^*, \btheta)} - 2 \sum_{i=1}^{n_{l-1}}\sum_{j=1}^{n_l} \log \prior(\hyper_{ij}^l).
\label{general:prior:integral3}
\end{aligned}
\end{equation}
Therefore we get the following cost function to be minimized in Eq~\eqref{general:proposition1} over ${\bw^l}, {\bG^l}, \btheta$:
\begin{equation}
\begin{aligned}
\mathcal{L}({\bw^l}, {\bG^l}, \btheta) 
=& {\bw^l}^{\top} \Hessian({\bw^l}^*, \btheta) {\bw^l}+ 2 {\bw^l}^{\top} \left[\Gradient({\bw^l}^*, \btheta)-\Hessian({\bw^l}^*, \btheta) {\bw^l}^{*}\right] +{\bw^l}^\top {\bG^l}^{-1} {\bw^l} 
\\
&+\log  |{\bG^l}|+\log |\Hessian({\bw^l}^*, \btheta)+ {\bG^l}^{-1} |
-N\log {{(2\pi\sigma^2)} \cdot b({\bw^l}^*, \btheta)} - 2 \sum_{i=1}^{n_{l-1}}\sum_{j=1}^{n_l} \log \prior(\hyper_{ij}^l).
\notag
\end{aligned}
\end{equation}
\end{pf}

\section{Proof for proposition~\ref{general:proposition2:cost}}
\label{app:proof for proposition 2}
\begin{pf}
	It can be easily found that Eq~\eqref{eq:general:cost:cccp} is consist of two parts, i.e. convex part in $\bw^l, \bG^l$ and concave part in $\bG^l$. For the first part, we define the function:
	{
		%	\small
		\fontsize{8.4}{2}
		\begin{equation}
		\begin{aligned}
		&{u}\left(\bw^l, \bG^l \right) 
		= {\bw^l}^{\top} {\Hessian^l} \bw+ 2{\bw^l}^{\top} \left[\Gradient^l-{\Hessian^l} {\bw^l}^{*}\right] +\bw^\top {\bG^l}^{-1} \bw^l \\
		\propto & (\bw^l-{\bw^l}^*)^{\top} {\Hessian^l} (\bw^l-{\bw^l}^*)+ 2{\bw^l}^{\top} \Gradient^l +\bw^\top {\bG^l}^{-1} \bw^l
		\label{general:summary:function:u}
		\end{aligned}
		\end{equation}
	}${u}\left(\bw^l, \bG^l \right)$ is a convex function jointly in $\bw^l$, $\bGamma^l$ as it is the sum of convex functions with type $f(\mathbf{x}, Y) = \mathbf{x}^{\top} \mathbf{Y}^{-1} \mathbf{x}$ \citep{boyd2004convex}.
	For the second part, we define the function:
	\begin{equation}
	\begin{aligned}
	v(\bGamma^l) & =  \log |\bGamma^l| + \log|{\bGamma^l}^{-1}  + {\Hessian}({\bw^l}^{*}, \btheta^*)|   \\
	& =   \log \left(|\bGamma^l|| {\bGamma^l}^{-1}  + {\Hessian}({\bw^l}^{*}, \btheta^*)| \right)  \\
	& = \log \left| \begin{pmatrix}
	{\Hessian}({\bw^l}^{*}, \btheta^*) & \\
	& -\bGamma^l
	\end{pmatrix} \right| \\
	& = \log\left|\bGamma +  {\Hessian}^{-1}({\bw^l}^{*}, \btheta^*) \right| + \log \left|{\Hessian}({\bw^l}^{*}, \btheta^*) \right|
	\label{eq:matrix determinant expansion}
	\end{aligned}
	\end{equation}
	$v(\bGamma^l)$ is a concave function since that Eq.~\eqref{eq:matrix determinant expansion} is a $\log$-determinant of an affine function of semidefinite matrices $\bGamma$.
	
	Therefore, the optimization problem in Eq.~\eqref{eq:general:cost:cccp} could be solved with a convex-concave procedure (CCCP). Specifically, by computing the gradient of concave part $v(\bGamma^l)$ to hyperparameter $\bgamma$, we have the following iterative optimization procedure:
	{
		%	\small
		\fontsize{9.5}{2}
		\begin{equation}
		\bw^l(t+1)
		=\argmin\limits_{\bw^l} u(\bw^l, \bgamma^l(t), {\Hessian}({\bw^l}^{*}, \btheta^*)) \label{eq:general:summary:eq:cccp1}
		\end{equation}
	}
	{
		%	\small
		\fontsize{9.5}{2}
		\begin{equation}
		\label{eq:general:summary:eq:cccp2}
		\begin{split}
		\bgamma^l(t+1)
		&=\argmin\limits_{\bgamma \succeq \mathbf{0}} u(\bw^l(t), \bgamma^l,{\Hessian}({\bw^l}^{*}, \btheta^*))+\nabla_{\bgamma^l} v(\bgamma^l(t), {\Hessian}({\bw^l}^{*}, \btheta^*))^\top\bgamma^l
		\end{split}
		\end{equation}
	}

\end{pf}
\section{Iterative Procedures to Update Parameters}
\label{appsec:cccp_procedure}
Eq~\eqref{eq:general:summary:eq:cccp2} provides the solution to update $\hat{{\bG^l}}$, the gradient of $v({\bG^l})$ to ${\bG^l}$ should be obtained firstly. By using the chain rule and the basic principle for convex analysis, its analytic form could be given as follows:
{
	\begin{equation}
	\begin{aligned}
	\balpha^l(t)
	\define &\nabla_{{\bG^l}} v\left({\bG^l},{\Hessian}({\bw^l}^*, \btheta^*)\right)^\top  |_{{\bG^l}={\bG^l}(t)}\\
	=&\nabla_{{\bG^l}}\left(\log |{\bG^l}^{-1}+{\Hessian}({\bw^l}^*, \btheta^*) |+\log  |{\bG^l}| \right)^\top|_{{\bG^l}={\bG^l}(t)}\\
	=& -\diag \left \{({\bG^l}(t))^{-1} \right \} \circ  \diag \left \{\left(({\bG^l}(t))^{-1}+{\Hessian}({\bw^l}^*, \btheta^*)\right)^{-1} \right \} \circ \diag \left \{({\bG^l}(t))^{-1} \right \} \\ & +\diag \left \{({\bG^l}(t))^{-1} \right \} \\
	=& 
	\left[
	\begin{array}{ccc}
	\alpha^l_{1}(t)& \cdots &  \alpha^l_{n_l}(t)
	\end{array}
	\right]
	\label{eq:general:alpha1k}
	\end{aligned}
	\end{equation}
}where $\circ$ represents the Hadamard product; where 
$\balpha^l(t)$ is the introduced intermediate variable and $\alpha^l_{j} = \left[\alpha^l_{1j}, \ldots, \alpha^l_{n_{l-1}j}\right]$. $\alpha^l_{ij}(t)$ could be obtained according to Eq~\eqref{eq:general:alpha1k}:
\begin{align}
\mathbf{C}^l(t)  &= \left({\bG^l}(t)^{-1}+{\Hessian}({\bw^l}^*, \btheta^*)\right)^{-1} 
\\
{\alpha_{ij}^l}(t) &= { -\frac{\mathbf{C}^l_{ij}(t)}{(\br^l_{ij}(t))^2} +\frac{1}{\br^l_{ij}(t)}} \label{eq:general:summary:gammastarupdate2}
\end{align}
Therefore, ${\bw^l}(t+1)$ and ${\bG^l}(t+1)$ can be calculated as iteratively as Eq~\eqref{eq:general:summary:eq:cccp1} and Eq~\eqref{eq:general:summary:eq:cccp2}:
{
	\begin{equation}
	\begin{aligned}
	& \left[{\bw^l}(t+1),{\bG^l}(t+1)\right]
	= \argmin\limits_{{\bw^l}} {\bw^l}^{\top} \Hessian^l {\bw^l} + 2{\bw^l}^{\top} \left(\Gradient^l -\Hessian^l {\bw^l}^{*}\right) 
	+\sum_{i=1}^{n_{l-1}}\sum_{j=1}^{n_l} \left(\frac{W^l_{ij}(t)^2}{\br^l_{ij}(t)} +\alpha^l_{ij}(t)\br^l_{ij}(t)\right).
	\label{general:summary:cccp-4}
	\end{aligned}
	\end{equation}
}
Since $$\frac{W^l_{ij}(t)^2}{\br^l_{ij}(t)} +\alpha^l_{ij}(t)\br^l_{ij}(t) \geq 2\left|\sqrt{\alpha^l_{ij}(t)} \cdot W^l_{ij}(t)\right|,$$ the optimal ${\bgamma^l}$ can be calculated as:
\begin{equation}
\begin{aligned}
\upsilon^l_{ij}(t)=\frac{|W^l_{ij}(t)|}{\sqrt{\alpha^l_{ij}(t)}},  \forall i,j.
\label{eq:general:summary:cccp-5}
\end{aligned}
\end{equation}
In order to compute ${\bgamma^l}(t+1)$, we also require the estimation for ${\bw^l}(t+1)$. 
If we define \begin{math}
\omega(t) \define \sqrt{\alpha(t)}
\end{math},
${\bw^l}(t+1)$ can be solved according to Eq~\eqref{general:summary:cccp-4}:
\begin{equation}
\begin{aligned}
{\bw^l}(t+1) =&\argmin\limits_{{\bw^l}}
\frac{1}{2}{\bw^l}^{\top} \Hessian^l {\bw^l}+ {\bw^l}^{\top} \left(\Gradient^l\ 
-\Hessian^l {\bw^l}^{*}\right) +\sum_{i=1}^{n_{l-1}}\sum_{j=1}^{n_{l}} \|\omega^l_{ij}(t) \cdot W^l_{ij}(t)\|_{\ell_1} \\
\propto&\argmin\limits_{\bw^l} E({\bw^l}^*, \btheta)+({\bw^l}-{\bw^l}^*)^{\top}\mathbf{G}({\bw^l}^*, \btheta)+\frac{1}{2}({\bw^l}-{\bw^l}^*)^{\top} \mathbf{H}({\bw^l}^*, \btheta) ({\bw^l}-{\bw^l}^*)  + \\ &2\sum_{i=1}^{n_{l-1}}\sum_{j=1}^{n_{l}} \|\omega^l_{ij}(t) \cdot \bw^l_{ij}\|_{\ell_1}
\\
\approx&\argmin\limits_{\bw^l} E(\cdot) + 2\sum_{i=1}^{n_{l-1}}\sum_{j=1}^{n_{l}} \|\omega^l_{ij}(t) \cdot \bw^l_{ij}\|_{\ell_1}
\label{eq:general:summary:rwglasso}
\end{aligned}
\end{equation}
then ${\bw^l}(t+1)$ could be injected into Eq~\eqref{eq:general:summary:cccp-5}:
\begin{equation}
\begin{aligned}
\br^l_{ij}(t+1)=\frac{|W^l_{ij}(t+1)|}{\omega^l_{ij}(t)}, \forall i,j.
\label{eq:general:summary:gammaik+1}
\end{aligned}
\end{equation}
With
Eq~\eqref{eq:general:summary:gammastarupdate2},~\eqref{eq:general:summary:rwglasso} and ~\eqref{eq:general:summary:gammaik+1},
the weight ${\bw^l}$ and hyperparameter ${\bG^l}$ could be updated alternatively. 
\end{sloppypar}
\end{document}